# Perturbation of Exciton Aggregate Coupling by Optical Excitation in Crystalline Perfluoropentacene Films

K. Kolata,[1] S. Marquardt,[2] N. W. Rosemann,[3] A. Wilm,[2] A.-K Hansmann,[2]
U. Höfer,[1] R. Berger,[2] T. Breuer,[1] G. Witte,[1] and S. Chatterjee,[1,3]

[1]Faculty of Physics and Materials Sciences Center, Philipps-Universität Marburg, Germany
[2]Faculty of Chemistry, Philipps-Universität Marburg, Germany
[3]Institute of Experimental Physics I, Justus-Liebig-University Giessen, Germany

## Abstact

Carrier multiplication by singlet exciton fission enhances photovoltaic conversion efficiencies in organic solids. This decay of one singlet exciton into two triplet states promises to overcome the Shockley-Queisser limit as up to two electrons may be harvested per absorbed photon. Intermolecular coupling is deemed mandatory for both, singlet exciton fission and a band-like transport. Such a coupling is manifested, *e.g.*, by the Davydov-splitting of the lowest-energy exciton transition in crystalline organic solids. For the model system perfluoropentacene, the corresponding transitions in the experimental, polarisation-resolved absorption spectra are identified by theoretical calculations based on the concept of H and J aggregation. Optical injection into the first vibronic progression of the fundamental exciton transitions significantly perturbs the higher-energy transitions that are associated to H-type aggregates of the $S_0 \rightarrow S_3$ transition during and following efficient singlet exciton fission. These findings underline the necessity for efficient carrier extraction as triplet accumulation may be detrimental to both, singlet exciton fission and any potentially band like transport. More generally, our observations indicate that electronic excitations can perturb the electronic band structure in organic crystals and highlight their correlated nature by potentially distorting the lattice.

## Introduction

Molecular materials are currently attracting significant attention in optoelectronics as cost-efficient alternatives to their inorganic counterparts.[1,2;3] Typical representatives of the latter are Si or GaAs used for detectors and emitters, respectively. These materials are crystalline allowing for a Bloch-function treatment of their optoelectronic response. Their emission spectra are typically dominated by energies associated with Wannier-type exciton states. These bound, Coulomb-correlated electron hole pairs typically extend beyond many unit cells of the crystalline lattice. The electronic excitations in such systems were intensely studied over the last decades,[4] *e.g.*, revealing the intricate interplay between excitons and free carriers.[5]

Generally, the optical properties of single molecules are qualitatively well described within the framework of transitions between different molecular orbitals when taking into account the appropriate selection rules and interactions such as Coulomb and exchange potentials. This infers, that excitations are thus confined to individual molecules in dilute systems and the lowest-energy dipole-allowed transition is typically observed between the highest occupied molecular orbital (HOMO) and the lowest unoccupied molecular orbital (LUMO). This optical absorption gap may be accompanied by vibrational progressions (often referred to as vibronic replica) and the fluorescence emission is then shifted to lower energies (bathochromic shift) according to the Franck-Condon principle.[6,7] Besides internal conversion, intersystem crossing into the triplet manifold of states by spin-flip may take place on longer time scales, where an electric dipole-forbidden transition to the ground state is possible via longer-lived phosphorescence. These triplet states are in extended unsaturated alternating hydrocarbon derivatives well shifted to lower energies due to large contributions of the exchange interaction.[8] Decreasing the intermolecular separation and studying less dilute systems potentially leads to excimer formation, where an excitation is shared between two adjacent molecules. Their hallmark is an additional bathochromic shift of the associated emission due to the energy gain of the system during intermolecular coupling.[9,10]

The molecular solid-state, however, is at the verge of these two well-established systems, the single molecule and the inorganic solid state. Excitations are generally still localized to some extent on the individual molecules forming the solid and hence may be described as Frenkel excitons.[11] They are manifested in additional absorption features, some of which at energies below the molecular optical absorption gap. Disordered systems may be treated using, *e.g.*, hopping models or by Dexter-Förster energy-transfer schemes due to the generally weak interaction between adjacent molecules.[12,13] Molecular crystals, however, show more pronounced, distinct signatures of intermolecular coupling. Probably most prominently, interactions within a bimolecular unit-cell feature a Davydov-splitting of exciton resonances.[11] This can go along with further delocalization of the excitations and charge transfer between adjacent molecules.[14,15] A long-range order may eventually lead to formation of dispersive electronic energy bands*, e.g.*, for favourable molecular orientations such as slip-stacking of delocalized π-electron systems.[16]

Additionally, slip-stacked systems have recently been shown to significantly enhance singlet-exciton fission (SF).[17,18] This unique feature of molecular materials describes the nonradiative conversion of one singlet exciton into two triplet excitons - the reverse process of triplet-triplet annihilation. This mechanism could double the number of excitations in a solar cell and hence should allow surpassing the Shockley-Queisser limit,[19] potentially constituting the prime physical advantage of organic photovoltaics. Consequently, quantum efficiencies beyond unity have been reported in model systems.[20] This process is expected for acenes with more than four fused benzene rings.[21] It is quite efficient in pentacene where the relative alignment of the singlet and triplet energy levels is favourable.[22,23] In this respect, also the role of vibron-assisted singlet-to-triplet exciton-conversion has been discussed.[24,25] Although the microscopic mechanism for SF is still debated in the literature, it is generally expected to be fostered by strong delocalization of excitations.

The detailed microscopic role of excitations and their effect on the electronic landscape remains unclear. In general, the fundamental transition energies in a molecular crystal are altered in the presence of excited carriers. After excitation, the system resides in a state with changed interaction potentials and, consequently, exhibits an absorption behaviour different from the linear optical response. Screening effects are less pronounced in organic semiconductors compared to their inorganic counterparts because of their different polarizability resulting in significantly larger exciton binding energies and the virtual absence of excitations as free carriers.[6,26] However, two effects should significantly alter the fundamental response: excitons localized on a given lattice site change the Coulomb potential of molecules in the vicinity, comparable to bandgap renormalization in inorganic semiconductors; secondly, excimer formation may break the initial symmetry of the crystal, either dynamically or due to relaxation effects inferring, that the Born-Oppenheimer approximation is no longer strictly valid.[6] This may lead to altered selection rules resulting in new, electric dipole allowed transitions.

K. Kolata et al.

Several approaches are commonly used to model the electronic properties of molecular solids. For example, density-functional theory with periodic boundary conditions may be applied in connection with GW and Bethe-Salpeter-equation-type modifications to identify the single-particle band structure and excitonic transitions.[27,28,29,30] These approaches, however, do not encompass vibronic coupling responsible for the dominant replica found in typical spectra of molecular solids. The latter are included in a molecular aggregation-type approach commonly applied in structurally less well-ordered systems or solution, which can also include crystalline symmetries.[31]

In order to investigate this role of the excited carriers in molecular systems we compare polarisation-resolved absorption spectra for various molecular arrangements and states of excitation by exploiting the relaxation dynamics following impulsive excitation. We distinguish three regimes: the linear response of the non-excited sample, a predominantly singlet-type exciton population during the first picoseconds, and the quasi-steady-state, long-term response by a population of diffused triplet-type excitons. We subject our findings to a detailed theoretical analysis taking into account vibronic interactions. Both singlet and triplet-type excitations perturb exciton aggregation signatures at energies well above the optical injection. This strongly suggests the need for efficient carrier harvesting to retain intermolecular coupling and enable the exploitation of the benefits of SF-enhanced solar converters.

## Results

**Signatures of H- and J-aggregates in the linear-optical response**

We begin by investigating signatures of intermolecular coupling in the polarisation-resolved linear optical response of perfluoropentacene[32] (PFP) films and, thereby, identifying the nature of the respective transitions. Preparing the molecular films on different substrates (NaF(001) and KCl(001)) yields crystalline films in different molecular orientation. Thereby, all molecular orientations can be directly accessed by optical spectroscopy in transmission geometry. Furthermore, since the crystalline domains are epitaxially aligned on these surfaces, an ideal polarization resolution is achieved.[33] The molecular arrangements in the crystalline thin-films are depicted in Fig. 1. On NaF (001), PFP forms domains in (100) orientation, which corresponds to an upright molecular configuration (*i.e.*, the molecular long axis, $\vec{L}$, is oriented perpendicular to the surface; although the molecules are slightly tilt towards the crystalline $\vec{c}$--direction by about 10°). This type of film growth enables spectroscopic access to the different packing motifs within the PFP crystal structure. In particular, the herringbone-packing motif along the $\vec{c}$-axis as well as the slip-stacked π-stacking arrangement along the $\vec{b}$-axis can be accessed individually to compare their polarisation-dependent absorption properties. A complementary orientation, *i.e.*, a lying molecular orientation is achieved on KCl(001). This allows a direct analysis of the optical properties associated with the molecular long axis.

The corresponding linear absorption spectra for light linearly polarised along the $\vec{c}$-axis and the $\vec{b}$-axis are shown in panels Fig. 1**c** and Fig. 1**d** as blue and red solid lines, respectively. The lowest-energy transitions in the solid-state are associated with a transition dipole moment oriented along the molecular short-axis $\vec{M}$.[34] A Davydov-splitting of about 27 meV is found. Each of the Davydov components is accompanied by a series of vibronic replica at higher energies. These bands closely resemble the lowest-energy electronic transition and its vibronic progression of PFP in solution, which is shown as grey-shaded area for comparison in all three panels. Consequently, they are sometimes also referred to in the literature as HOMO-LUMO transition in the solid state with its vibronic progressions.^Fehler! Textmarke nicht definiert. For light which is linearly polarisation along the $\vec{a}$-axis, only residual signatures are found in this energy range since the short molecular axis is not efficiently excitable in this case.

Next, we turn to higher energies. Notably, we find a second band of transitions in the range from 2.3 to 3 eV for the polarisation along the $\vec{c}$-axis which superimposes the lowest-energy excitons' vibronic progressions. This higher-energy absorption band ivirtually absent when the probe light polarisation is set parallel to the $\vec{b}$-axis. By comparing solid-state absorption with the solution absorption spectrum, strikingly, the energy spacing between the progressions is increased from ~180 meV (~1450 cm$^{-1}$ $hc$) in solution to ~190 meV (~1530 cm$^{-1}$ $hc$) in the solid state. This could imply a concomitant upshift in vibrational frequency, a more restricted change in equilibrium structure upon electronic excitation that renders progressions in a different normal mode more prominent or an additional splitting induced by excitonic coupling. The dominant resonance in this energy range is observed along the $\vec{a}^*$-axis.[33,34]

**Modelling of H and J aggregates**

To assign the higher-energy absorption and, in particular, to physically understand the anisotropic optical response, we perform quantum chemical calculations on isolated molecules and subsequently use a model Hamiltonian description of the excitonic coupling (see section computational details for an in-depth description). The vertical excitation energies computed on the density functional theory (DFT) level at the equilibrium structure of the electronic ground state reveal thatthe transition dipole moment of the HOMO → LUMO transition ($S_0$ → $S_1$),is polarised along the short molecular axis ($\vec{M}$) (Fig. 1**c**), three additional electronic transitions appear in this energy range, two of which are electric dipole forbidden in the Franck-Condon approximation (cf. Table 1). Only the $S_0$ → $S_3$ transition (HOMO → (LUMO+1)) is electric dipole allowed, with the polarisation being aligned with the long molecular axis ($\vec{L}$). These high-energy features are attributed to a different fundamental electronic transition ($S_0$ → $S_3$) than the absorption bands at lower energies, which correspond to the $S_0$ → $S_1$ transition as will be discussed in the following section in detail.

We determine the Franck-Condon-profiles from the computed equilibrium structures, the normal modes, and the harmonic vibrational wavenumbers of the isolated molecules, *i.e.*, the vibronic spectra for an electronic excitation from the electronic ground state to different electronically excited singlet states ($S_0$ → $S_1$, $S_0$ → $S_3$) for temperatures of 300 K. These profiles agree reasonably well with the measured solution spectra as shown in Fig. 2**a**). Note that in the graphical representation, the relative intensities and energies of the 0-0 transitions are adjusted to the experimental spectra. We estimate an effective Huang-Rhys-factor (related to $d_f$, see section on computational details) from the multimode Franck-Condon profiles for a subsequent effective single-mode description in the various electronic states; this factor is commonly used to describe the coupling of vibronic and electronic excitations through the difference between ground-state and excited-state geometries.

For interpretation of the main features in the spectra measured for the crystalline film, we follow essentially the theoretical model in Ref. [35] for ideal J- and H-aggregate absorptions. In this theoretical modelling, the absorption behaviour is analysed for varying electric dipole-coupling strength taking into account vibronic coupling as determined by the Huang-Rhys factor from the single molecule calculations.[36] Together with the excitonic coupling parameter *L*, which is estimated initially based on the electric transition dipole moments in combination with the molecular arrangement and later derived from the experimental data, we calculate the linear absorption spectrum for the solid state.

**Table 1:** Energies, oscillator strengths, and electric transition dipole moment orientations of PFP absorption lines in the gas phase on the density functional theory level. See computational details for methods. The axis convention for assignment of the irreducible representation was x along the long molecular axis and y along the short molecular axis.

| Vertical excitation energy ($E_{ex}$) (in eV) | Irreducible representation ($\Gamma$) | Oscillator strength ($f$) | Transition dipole moment component along molecular $\vec{L}$-axis ($\mu_L$) in units of $e\,a_0$ | Transition dipole moment component along molecular $\vec{M}$-axis ($\mu_M$) in units of $e\,a_0$ | Transition dipole moment component along molecular $\vec{N}$-axis ($\mu_N$) in units of $e\,a_0$ | Transition dipole moment ($|\vec{\mu}|$) in units of $e\,a_0$ | Nature of transition |
|---|---|---|---|---|---|---|---|
| 1.68 | $b_{2u}$ | 0.039 | 0.00 | 0.97 | 0 | 0.97 | HOMO → LUMO |
| 2.65 | $b_{1g}$ | 0.000 | 0.00 | 0.00 | 0 | 0.00 | (HOMO-1) → LUMO |
| 2.96 | $b_{3u}$ | 0.316 | 2.09 | 0.00 | 0 | 2.09 | HOMO → (LUMO+1) |
| 3.06 | $b_{1g}$ | 0.000 | 0.00 | 0.00 | 0 | 0.00 | HOMO → (LUMO+2) |
| 3.63 | $b_{2u}$ | 0.000 | 0.00 | -0.06 | 0 | 0.06 | (HOMO-2) → LUMO |
| 4.00 | $a_g$ | 0.000 | 0.00 | 0.00 | 0 | 0.00 | (HOMO-1) → (LUMO+1) |
| 4.18 | $b_{3u}$ | 2.681 | -5.12 | 0.00 | 0 | 5.12 | (HOMO-3) → LUMO |

The results are shown in Fig. 2**b** and **2c**. The corresponding calculated monomer spectrum (single molecule absorption) is given as a grey shaded area for reference. The energy shift of the transition resulting from the solid-state background is omitted in order to emphasize the effects of aggregation. In Fig. 2**b,** we show the results for a one-dimensional molecular chain with arbitrary parameters. This chain model is motivated by the packing motif of the slipped $\pi$-stacking along the $\vec{b}$-axis. The blue curve shows the absorption spectrum of a pure J-aggregate. In this case the transition dipoles are all directed parallel to the molecule chain with head-to-tail or tail-to-tail arrangement. Here, the vibronic progressions appear to besuppressed and the lowest-energy transition gains oscillator strength. Furthermore, the overall absorption is shifted towards lower energies. The H-aggregate shown in red, for which the transition dipoles are orthogonal to the chain direction and parallel to each other, shows opposing effects. The intensity of the 0-0-transition is decreased and the intensity is shifted towards the region of the vibronic progressions. The curves in purple and orange correspond to the two possible electronic transitions for a system with oblique transition dipole moments. An absorption with light polarised in the orthogonal direction to the chain shows H-aggregate-like behaviour. In the case when the light is polarised in the direction of the chain the absorption is J-aggregate-like.

Figure 2**c** shows the linear absorption spectra with parameters derived from the theoretical data of the S$_0$ → S$_1$ transition. The coupling motifs between the transition dipole moments aligned with the short molecular axis ($\vec{M}$) are essentially confined to the $\overrightarrow{(b,c)}$-plane, whereas coupling along the $\vec{a}$-axis is reduced. This motivates the use of a two-dimensional model for the description of the excitons in this case. Here, we choose an effective tetramer model (pinwheel) for simplicity to highlight the main features. Thus, in Fig. 2**c** we use coupling

K. Kolata et al.strengths that are larger than computed for the molecular dimer. For all three polarization directions, we observe the J-aggregate-like shift that is also found in the experiment. We also find that with polarisation along the $\vec{a}^*$-axis the intensities are much smaller than along the other axes.

We performed similar calculations also for the $S_0 \rightarrow S_3$ transition, whose electric transition dipole moment is aligned with the long molecular axis ($\vec{L}$). The results are shown in Fig. 2**d**. The spectra for polarisation along the $\vec{b}$-axis and along the $\vec{c}$-axis are scaled by factors of 10000 and of 10, respectively. The curves follow right trends, along the $\vec{a}^*$-axis, the absorption is shifted to higher energies whereas along the $\vec{b}$-axis and the $\vec{c}$-axis, the absorption is shifted to lower energies. The intensities for polarisation along the $\vec{c}$-axis are higher than for polarisation along the $\vec{b}$-axis, which is consistent with the molecular orientation. The molecules are tilted more strongly towards the $\vec{c}$ direction, such that the projection of the electric transition dipole moment on the $\vec{c}$-axis is larger than on the $\vec{b}$-axis.

Taking into account all the above considerations yields the following association of the peaks: peak I-IV are attributed to excitonic transitions related to the molecular $S_0 \rightarrow S_3$ transition as this features a significant dipole moment. An attribution to the $S_0 \rightarrow S_1$ transition appears unreasonable; this has lower oscillator strength, which implies relatively weak excitonic coupling. This, however, contradicts the vast energy shift observed. The $S_0 \rightarrow S_1$ transition appears at lower energies for all employed polarisation directions.

**Polarisation-resolved pump-probe experiments**

Next, we analyse the influence of excited carriers on the stronger intermolecular coupling along the $\vec{b}$-axis by polarisation-resolved pump-probe experiments. Here, we focus on the absorption-band located in the energy region between 2.2 and 2.6 eV to study the influence of excitation on intermolecular coupling whereas the response at the lower-energy resonances around 1.7 eV is indicative for blocking of transitions and occupation of the excitonic states themselves. The relevant changes of the linear absorption due to excitations in the system in this energy range are summarized in Fig. 3. Here, the differential absorption spectra, *i.e.*, the pump induced change in the absorption, $\Delta\alpha L$, are given in false colours as function of the probe photon-energy (horizontal axis) and pump-probe time delay (vertical axis). The probe polarisations are set parallel to the crystal's $\vec{b}$ axis in Fig. 3**a** and along its $\vec{c}$ axis in Fig. 3**b**; the pump polarisation is set along the $\vec{b}$ axis in both cases. Negative differential absorption signals are encoded in blue, whereas positive signals are displayed in green in the region of interest. The low-energy signals in the vicinity of the lowest singlet-exciton transition and its first phonon replica are greyed out. Here, we find the characteristic bleaching accompanied with a slight shift, which is manifested by an induced-absorption-like signature at energies above the transition energy.[37] Clearly, the non-linear responses are highly anisotropic. Furthermore, two distinct time-regimes are identified marked by the olive and purple dashed boxes.

To better quantify the observed changes and hence identify their origin, averaged differential absorption spectra for both axes are plotted in Fig.2**c** and Fig.2**d**. The olive curves show the average $\Delta\alpha L$-signal for time delays from 0-12 ps and the purple curves span time delays from 12-1000 ps. These two time regimes capture two different excitation conditions: shortly after excitation the excitons reside in singlet-type, excimeric or correlated triplet pair states; singlet fission occurs on this timescale. Hence, the excitations reside in diffused triplet-type states at later times. [38] The system is in a quasi-equilibrium that slowly decays by recombination of

6 / 17

triplet-type excitons. Again, the lower-energy responses are greyed out and they are scaled in height for clarity, the corresponding factors are provided in the figure.

All features along the $\vec{c}$-axis (Fig 3**d**) in the high-energy range are strongly bleached.[38] When comparing the two time regimes depicted in Fig. 3**d**, no shifts in energy are observed. The energy difference between the two bleaching peaks at 2.29 eV and 2.49 eV remains constant for singlet-type and triplet-type excitations in the system associated with early and late time delays, respectively. Hence, these signatures are interpreted as direct results of a *reduced* J-aggregate-like absorption. Consequently, the intermolecular dipole coupling during optical excitation is *decreased* due to the presence of an excitonic population. The remaining induced absorption results from the shift signature associated with the fundamental singlet exciton transition and its vibronic replica.

The pump-probe spectra recorded parallel to the $\vec{b}$-axis show a distinctively different response (Fig 3**c**). Several induced absorption features appear instead of the previously discussed bleaching observed for polarisation of the probe pulse along the $\vec{c}$-axis. Additionally, the features evolve spectrally with time, slightly shifting to lower energies. As far as the line-shape is concerned, the peaks resemble vibronic progressions like the energy band starting at 2.29 eV in the linear absorption along the $\vec{c}$-axis discussed above. The emerging progressions observed along the $\vec{b}$-axis thus resemble the linear absorption of the $\vec{c}$-axis. As this was attributed to an H-aggregate of the molecular long-axis dipole, which should not be observed in the equilibrium lattice structure. This may hint towards an excitation-induced slight crystalline deformation during the initial few ps that enables addressing this transition by light polarized along the $\vec{b}$-axis. Alternatively, this signature may be similar to the solution spectra, *i.e.*, as if monomer-like uncoupled molecular dipole transitions become increasingly allowed when excitons are present in the system. Another change in the background potentials is observed once the singlet-like excitons diffuse into individual triplet excitons during the fission process, *i.e.*, the correlated triplet pairs $^1(TT)$, after approximately 12 ps (purple solid line in Fig. 3**d**). Now, the vibronic progressions shift to lower energies and the relative oscillator strength of the individual contributions is modified.

**Discussion**

Molecular materials offer intriguing properties such as singlet exciton fission, which may allow surpassing the Shockley-Queisser limit for photovoltaic devices. Single crystals are ideal model systems to explore the microscopic origin and clearly identify the nature of the required intermolecular coupling. The optical response of these materials commonly bears significant contributions of the vibronic structure. These are included in an exciton aggregate model which enables the assignment of several excitonic resonances to the polarisation-resolved linear absorption spectra of crystalline PFP films. In this work, we study the anisotropic linear absorption of crystalline PFP to identify all solid-state absorption bands in the visible spectrum and assign their origin as compared to the features observed in solution spectra. Our observations are discussed within an exciton-aggregate model taking into account the Huang-Rhys coupling parameters within this system.as well as by DFT-based calculations of their respective influence on the optical absorption bands.

Based on these results, the effects of optically injected excitons on the intermolecular coupling of electric transition dipole moments are explored. The experimental observations are theoretically analysed within the framework of J-aggregates and H-aggregates: the optical injection of carriers leads to a reduction of signatures that are attributed to intermolecular

K. Kolata et al.

coupling. Therefore, the molecular coupling is decreased by the presence of electronic excitations in the system. Any kind of excitation, *i.e.*, singlet or triplet-type excitons, excimers, and correlated-triplet-pairs, $^1(TT)$, perturb the periodicity required to observe a collective dipole coupled response and signatures of uncoupled individual molecules prevail. Hence, all excitation suppress the collective dipole-coupled response that would ultimately lead to a band-like transport. Alternatively, a slight deformation of the crystal lattice by the optical excitation is expected due to the appearance of additional resonance in the polarisation-resolved induced absorption. Thus, efficient carrier extraction is required after singlet fission to prevent an accumulation of triplet excitons, which potentially reduces the efficiency of additional singlet-exciton fission processes.

## Methods

**Sample Growth and Structural Characterization:**

The PFP (Kanto Denka Kogoyo, purity > 99%) films are grown on crystalline NaF (001) and KCl (001) surfaces. The alkali halide surfaces are prepared by cleaving slices of about 2 mm thickness from a single-crystal rod (Korth Kristalle GmbH) in air. After transfer into the vacuum system, the substrates are annealed at 450 K to remove adsorbed water. Subsequently, the highly-crystalline PFP thin films (150 nm) are prepared under ultra-high-vacuum conditions by molecular beam deposition at a molecular flux of about 6 Å/min as monitored by a quartz crystal microbalance. To maximize the domain sizes, the PFP films are grown at a substrate temperature of 350K. PFP forms epitaxially ordered adlayers of the bulk structure.[33] On NaF substrates, the molecules adopt an upright (100)-orientation with lateral alignment along the substrate <100> directions. On KCl, the PFP molecules are arranged in the <102>-orientation which corresponds to a lying molecular geometry. In this case, the crystalline domains are aligned along the KCl <110> directions. The crystalline structure of all samples is verified by X-ray diffraction, optical microscopy, and atomic force microscopy as detailed in 33.

**Pump-probe spectroscopy and data evaluation**

All polarisation-resolved pump-probe experiments are performed at room temperature. The laser source driving both, the pump and the probe beam is a regenerative 100 kHz Ti:sapphire amplifier system. The samples are excited by strong pump pulses generated in an optical parametric amplifier tuned to 1.95 eV. The photon flux is set to $4.5 \times 10^{15}$ photons/cm$^2$ per pulse. A weak fs-white-light supercontinuum is used as probe. It is generated in an yttrium-aluminium-garnet (YAG) crystal using a small fraction of the fundamental 800 nm pulse intensity. The linear polarisation of both arms, i.e., pump and probe, is defined by combinations of two Glan-laser-type polarizers and broadband half-wave plates adjusted to the desired angle with respect to the plane of incidence before they are focused onto the sample. This way, a polarisation contrast better than 1000:1 is ensured in both cases. Furthermore, the angle of incidence of the pump pulse is kept small, so that out of plane excitations are negligible. The setup yielded sub-300 fs time resolution.

The linear absorption at the excitation energy 1.95 eV is nearly degenerate for the $\vec{b}$- and the $\vec{c}$-axis. Therefore, for all experiments, the pump pulse is kept linearly polarised along the $\vec{b}$-axis. The probe polarisation is adjusted by checking the spectral shift of the fundamental transition in the transmission spectrum. The maximum blue shift in this transition corresponds to the $\vec{b}$-axis response of the crystal. Moreover, the amount of shift as well as the linewidth of the fundamental transition are very good indicators for the domain quality. The orientation of the probe pulse polarisation allows for the correlation of structural properties and electronic excitation. Its polarisation is thus set parallel to the $\vec{b}$- and $\vec{c}$-axis.

After passing through the sample, the probe is dispersed in a spectrometer with a spectral resolution of 1 nm and is detected by a thermoelectrically cooled silicon charged coupled device camera (1064x120 pixels) cooled to -30°C. The change of absorption is measured by opening and closing pump and probe arm with mechanical shutters.

The $\Delta\alpha L$ is calculated as follows:

$$\Delta\alpha L = -\ln\left(\frac{T_{\text{PPr}} - T_{\text{P}}}{T_{\text{Pr}} - T_{\text{B}}}\right)$$

$T_{\text{PPr}}$ is the transmission of both, pump and probe. $T_{\text{P}}$ captures only the pump transmission with the probe arm closed in order to correct for scattered pump light and photoluminescence from the sample. $T_{\text{Pr}}$ is the probe transmission corrected by the background $T_{\text{B}}$ (both shutter closed). Hence, the experiment is able to monitor the pump-induced changes in the absorption of the sample as a function of polarisation, probe photon energy, and time.

**Computational details**

Gas phase vibronic spectra of perfluoropentacene monomers are computed in the adiabatic and harmonic approximation with the program hotFCHT [39,40,41,42]. Equilibrium structures on the Born-Oppenheimer hypersurface together with the harmonic vibrational force fields in the electronic ground state and the energetically lowest excited singlet state are computed on the level of Kohn-Sham density functional theory (DFT) [43] or its time-dependent variant (TD-DFT) [44] with the def2-TZVP basis set [45,46] and the B3LYP functional [47,48] with the program package Turbomole [49]. The electronic ground state, the HOMO-LUMO excited state and the HOMO-(LUMO+1) excited state are found to have $D_{2h}$ symmetric equilibrium structures. This point group symmetry is exploited when computing the harmonic vibrational force fields. Excitonic coupling is described with an effective Hamiltonian that allows for static excitonic coupling between neighbours and considers only a single effective linear electron-phonon coupling mode. By virtue of the computed gas phase vibronic spectra for the $S_0 \to S_1$ transition and for the $S_0 \to S_3$ transition, which accounts for linear and all quadratic coupling terms (full Duschinsky mode mixing is included), the harmonic vibrational wavenumber $\widetilde{\omega}_e$ of the single effective mode is chosen ($\omega = 2\pi c \widetilde{\omega}_e$). The effective linear vibronic coupling strength $d_F$ is determined by the intensity ratio between the two spectral regions of lowest wavenumbers (typically from 0 cm$^{-1}$ to 1000 cm$^{-1}$ and from 1000 cm$^{-1}$ to 2000 cm$^{-1}$ with respect to the 0-0 transition wavenumber). The Frenkel exciton coupling strength $L$ was initially chosen to resemble the interaction energy between the respective electric transition dipole moments when located at the two different molecules in the unit cell and later adjusted. These parameters are listed in Table 2.

This model bases on the frequently used dimer model of excitonic coupling as described in Refs. [50,51] (see also Ref.[52] for a recent review ) and related models for oligomers.[53] In all our calculations we use open boundary conditions together with a many-body expansion of the excitonic wave function, which was terminated at the two-body level. Vibrational excitations up to quantum numbers $v = 4$ and $v' = 4$ in the electronic ground state and electronically excited state of each monomer are included (parameter $v_{\max}$). The effective Hamiltonian used reads in second quantized language as $\widehat{H}_{\text{eff}} = \widehat{H}_{\text{ex}} + \widehat{H}_{\text{vib}} + \widehat{H}_{\text{ex-vib}}$ with the Frenkel exciton part $\widehat{H}_{\text{ex}} = \sum_i E_F \hat{A}_i^+ \hat{A}_i + \sum_{i,k} L(\delta_{i,k+1} + \delta_{i,k-1})\hat{A}_i^+ \hat{A}_k$ the phonon vibrational part $\widehat{H}_{\text{vib}} = \sum_i \hbar\omega \; (\hat{a}_i^+ \hat{a}_i + \frac{1}{2})$ for the single vibrational degree of freedom at each site $i$. Here, the energies $E_F$ of the uncoupled single site excitons and $\hbar\omega$ of the effective single mode phonons as well as the excitonic coupling strengths $L$ and the linear vibronic coupling strength $d_F$ given below are parameter of the model. The operator $\hat{A}_i^+$ creates an exciton at site $i$, whereas $\hat{A}_i$ annihilates a corresponding exciton. Similarly, $\hat{a}_i^+ (\hat{a}_i)$ creates (annihilates) a phonon at site $i$. The linear electron phonon coupling term reads as $\widehat{H}_{\text{ex-ph}} = \sum_i \hat{A}_i^+ \hat{A}_i \, d_F(\hat{a}_i^+ + \hat{a}_i) + d_F^2/(\hbar\omega)$. We use in our implementation a unitarily transformed description in a displaced, undistorted harmonic-oscillator basis for the excitonic state and compute matrix elements between the undisplaced harmonic oscillator in the electronic ground state and the displaced harmonic oscillator in the electronically excited (excitonic) state. On each site, we allow for phonon excitation up to the quantum numbers mentioned above. Only one-body and two-body terms are kept in the expansion of the full wave function which means that only basis functions with up to single excitons and up to phonons excited at two different sites are included. To describe absorption features, electromagnetic coupling of linearly polarized light to the electric transition dipole moment operator at each site is considered and a Gaussian line width function with a full width at half maximum of $0.3 \; \hbar\omega$ was assumed.

**Table 2:** Parameters for the simulations of absorption spectra. The signs where given in parenthesis were chosen differently for the different polarisation directions depending on either the projection of the transition dipoles on the polarisation axis showed an H-aggregate or J-aggregate-like arrangement.

|  | Linear chain | Linear chain | Linear chain | Monomer | Pinwheel | Pinwheel |
|---|---|---|---|---|---|---|
|  | J-aggregate | H-aggregate | Tilted dipoles |  | $S_0 \to S_1$ | $S_0 \to S_3$ |
| $\widetilde{\omega}_e$ | 1400 cm$^{-1}$ | 1400 cm$^{-1}$ | 1400 cm$^{-1}$ | 1400 cm$^{-1}$ | 1400 cm$^{-1}$ | 1200 cm$^{-1}$ |
| $d_F$ | 1 $\hbar\omega$ | 1 $\hbar\omega$ | 1 $\hbar\omega$ | 1 $\hbar\omega$ | 0.9 $\hbar\omega$ | 0.71 $\hbar\omega$ |
| $L$ | $-0.5 \, \hbar\omega$ | $0.5 \, \hbar\omega$ | $(-)0.5 \, \hbar\omega$ |  | $(-)0.1\hbar\omega^*$ and $(-)0.5\hbar\omega$ | $(-)0.45\hbar\omega^*$ and $(-)2.0\hbar\omega$ |
| Number of molecules | 5 | 5 | 5 | 1 | (2x2) | (2x2) |
| $\nu_{\text{max}}$ | 4 | 4 | 4 | 4 | 4 | 4 |

*Not shown, calculated coupling for dimer

**Data availability**

Data supporting the findings of this study are available within the article (and its Supplementary Information files) and from the corresponding author on reasonable request.

**Acknowledgments:**

We thank J. Güdde for help with the regenerative Ti:Sapphire amplifier system. Financial support is provided by the German Science Foundation (DFG) through the collaborative research center "Structure and Dynamics of Internal Interfaces" (SFB 1083).

**Author contributions:**

S.C. and G.W. conceived and designed the experiments; K. K., N.W.R. and S. C. performed the optical spectroscopy and interpreted the data; T. B. and G. W. prepared and structurally investigated the samples; R.B. applied the theory and the model; A. W. and S. M. implemented the effective excitonic coupling Hamiltonian, A.-K H. computed the required electronic structure data, the harmonic vibrational frequencies, the vibronic spectra of PFP and the various singlet excitation energies; all authors co-wrote the paper.

**Additional information:**

Supplementary information is available in the online version of the paper. Reprints and permissions information is available online at www.nature.com/reprints. Correspondence and requests for materials should be addressed to sangam.chatterjee@physik.uni-giessen.de

**Competing financial interests:**

The authors declare no competing financial interests.

K. Kolata et al.

**Figure 1**

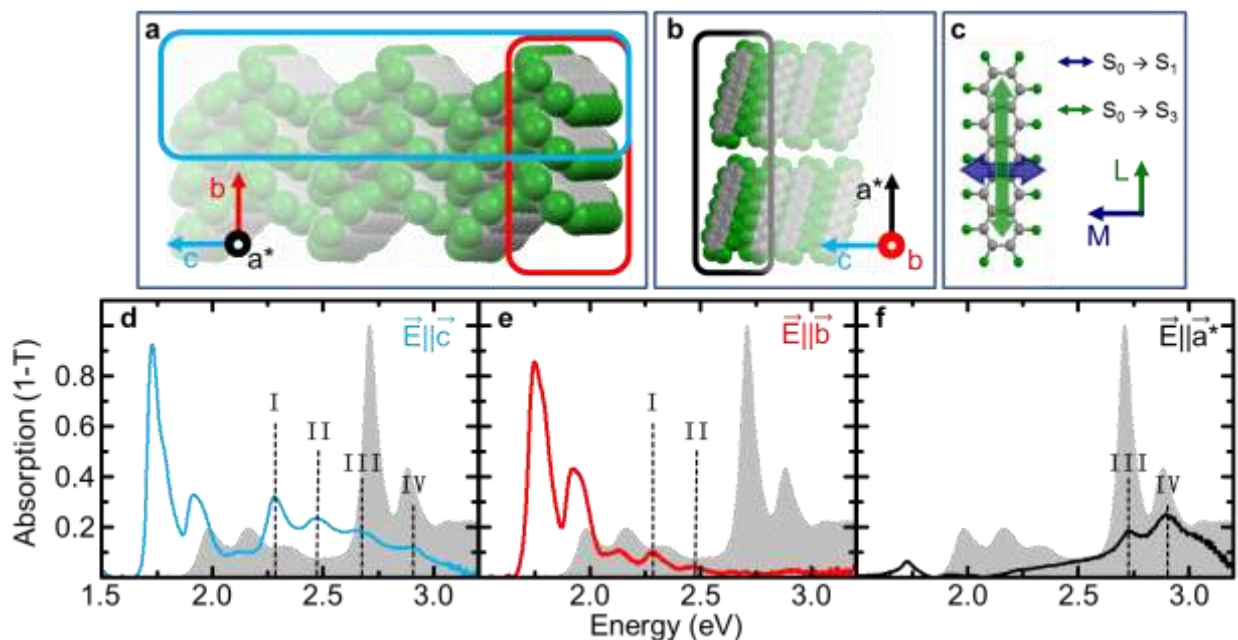

**Figure 1: Overview on the crystal structure and linear optical absorption. a,b)** Visualization of molecular packing as adopted in the investigated PFP films on NaF (001) in top-view (**a**) and side view (**b**), **c**) molecular geometry of PFP molecule with orientation of dipole moments associated to $S_0 \rightarrow S_1$ and $S_0 \rightarrow S_3$ transitions. **d-f**) Linear absorption spectra of PFP thin films acquired with light polarizations along different crystallographic directions: **d**) E ∥ $\vec{c}$, e) E ∥ $\vec{b}$, f) E ∥ $\vec{a}$. For comparison, a PFP spectrum in dichloromethane solution is provided as grey shaded curve (data according to 34}. (**d**) and **e**) were acquired from measurements of PFP films on NaF (001), while **f**) was acquired from PFP films prepared on KCl(001).)

K. Kolata et al.

**Figure 2**

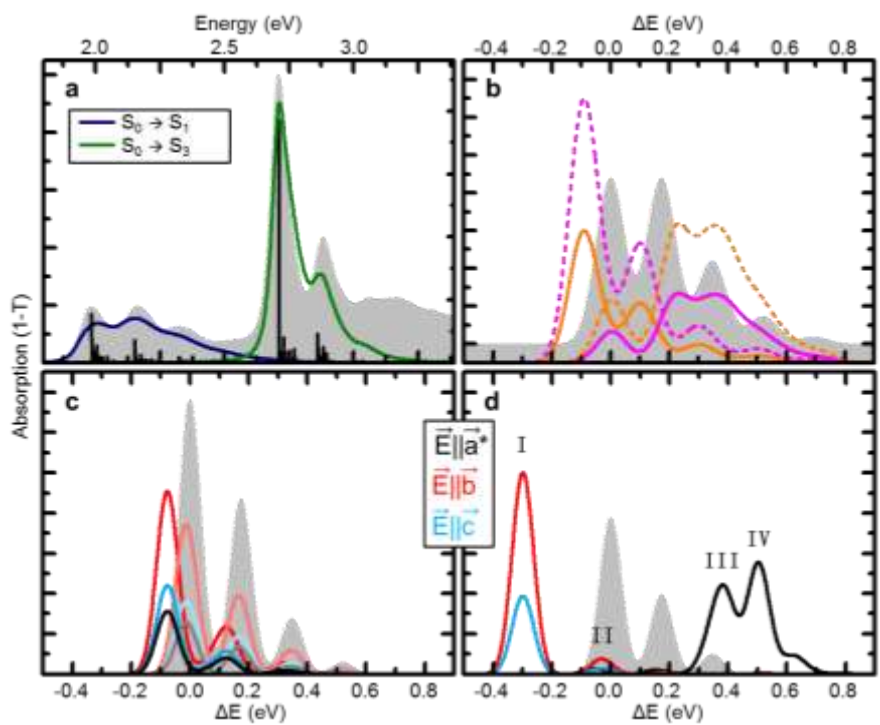

**Figure 2: Polarisation-resolved absorption spectra including monomer and solution spectra (grey, for reference).** (**a**) Calculated Franck-Condon profiles. The spectra at 0 K are shown as stick spectra. The spectra at 300 K are shown in blue ($S_0 \rightarrow S_1$) and green ($S_0 \rightarrow S_3$). The signals are shifted and scaled in order to match the experimental spectrum. (**b**) Calculated linear absorption spectra for a linear chain as a J-aggregate (pink), an H-aggregate (orange) or with oblique transition dipoles (short dashed). (**c**) Calculated linear absorption spectra for a (2x2)-system resembling the $S_0 \rightarrow S_1$ transition. The results are for polarisation along the a-axis (black), the b-axis (red) and the c-axis (blue). (**d**): Calculated absorption spectra for a (2x2)-system resembling the $S_0 \rightarrow S_3$ transition. The colour code is the same as for subfigure c. The signals for polarisation along the b axis and the c axis are scaled by factors of 10000 and 10, respectively.

K. Kolata et al.

**Figure 3**

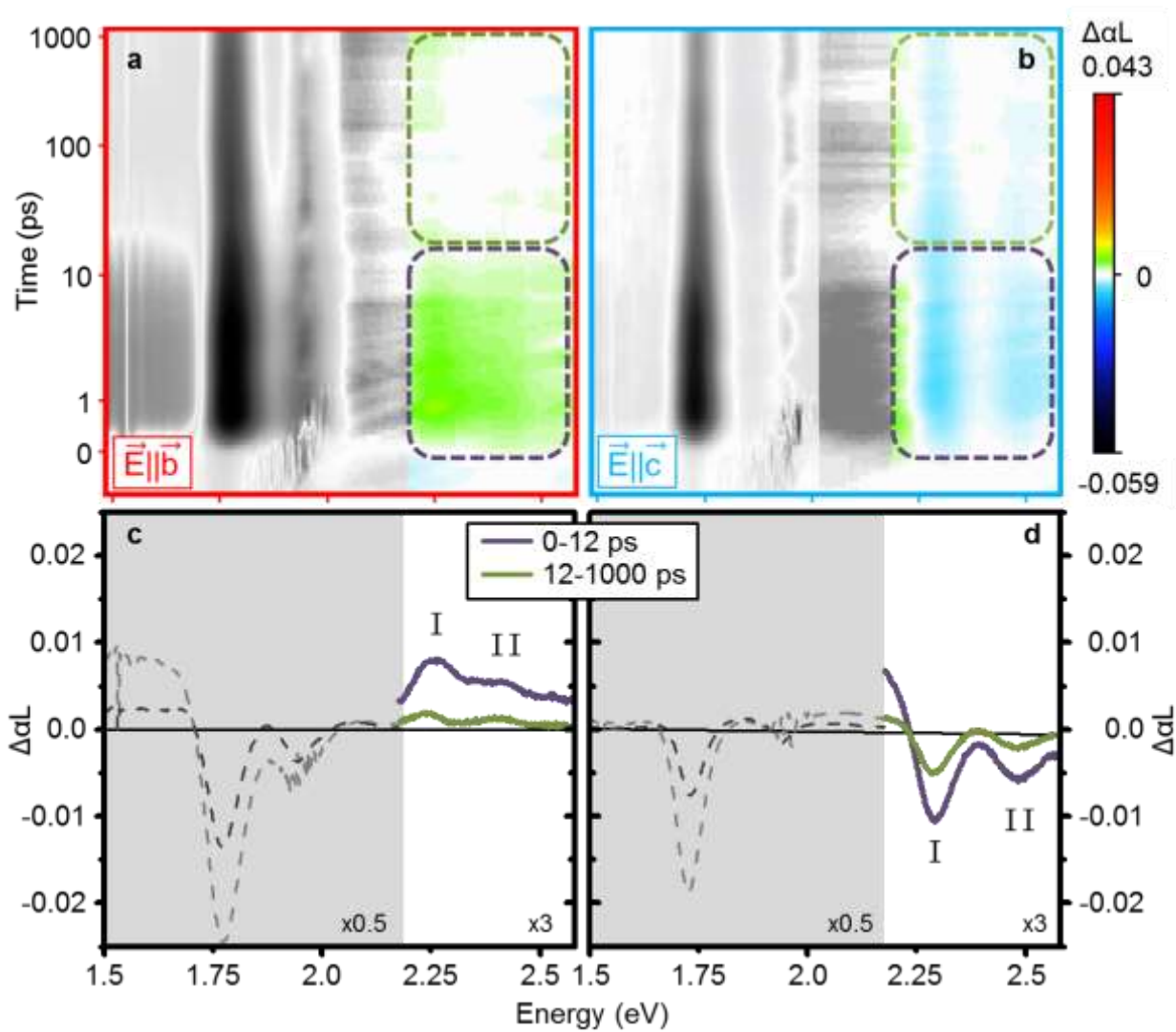

**Figure 3: Polarisation-resolved transient absorption spectroscopy. a,b)** Time-dependent differential absorption spectra of PFP films with different light polarizations of the probe source revealing strongly anisotropic behaviour in the energetic regime between 2.2 eV and 2.6 eV. The purple and olive boxes, respectively, highlight the time regions, which are used to extract the spectra shown in **c**) and **d**). Significant induced absorption at the positions of signals I and II is observed for polarization along the $\vec{b}$-axis, while a bleaching of these signals is found along the $\vec{c}$-axis. Both effects are strongest in the early time regime (0-12 ps) but still observed after up to 1 ns. (Note that the strong bleaching observed at lower energies is a direct effect of the increased occupation initiated by the pump-pulse and therefore not discussed in detail here.)

K. Kolata et al.

# References

1. Piner, R. D., Xu, J. Z., Hong, S., Mirkin, C. A. "Dip-pen" nanolithography. *Science* **283,** 661–663 (1999).
2. Sirringhaus, H. *et al*. High-resolution inkjet printing of all-polymer transistor circuits. *Science* **290,** 2123–2126 (2000).
3. Krebs, F. C., Gevorgyan, S. A., Alstrup, J. A. Roll-to-Roll Process to Flexible Polymer Solar Cells: Model Studies, Manufacture and Operational Stability Studies. *J. Mater. Chem.* **19,** 5442-5451 (2009).
4. Shah, J. *Ultrafast Spectroscopy of Semiconductors and Semiconductor Nanostructures,* Springer (1999).
5. Chatterjee, S. *et al*. Excitonic photoluminescence in semiconductor quantum wells: plasma versus excitons. *Phys. Rev. Lett.* **92**, 067402 (2004).
6. Schwoerer, M., Wolf, H.C. *Organic Molecular Solids* Wiley (2007).
7. Clark, J., Silva, C., Friend, R. H., Spano, F. C. Role of intermolecular coupling in the photophysics of disordered organic semiconductors: aggregate emission in regioregular polythiophene. *Phys. Rev. Lett.* **98,** 206406 (2007).
8. Köhler, A, Bässler, H. Triplet states in organic semiconductors. *Mater. Sci. Eng. R-Rep*. **66,** 71-109 (2009).
9. Shirai, S.; Iwata, S.; Tani, T.; Inagaki, S. Ab initio studies of aromatic excimers using multiconfiguration quasi-Degenerate Perturbation Theory. *J. Phys. Chem. A* **115,** 7687–7699 (2011).
10. Beljonne, D.; Yamagata, H.; Brédas, J. L.; Spano, F. C.; Olivier, Y. Charge-Transfer Excitations Steer the Davydov Splitting and Mediate Singlet Exciton Fission in Pentacene. *Phys. Rev. Lett.* **110,** 226402 (2013).
11. Davydov, A. S. The theory of molecular excitons. *Sov. Phys. Usp.* **7,** 145 (1964).
12. Olaya-Castro, A., Scholes, G.D. Energy transfer from Forster-Dexter theory to quantum coherent light-harvesting, *Inter. Rev. Phys. Chem.* **30,** 49-77 (2011)).
13. Frenkel, J. On the transformation of light into heat in solids. II. *Phys. Rev.* **37,** 1276-1294 (1931).
14. Sharifzadeh, S., Biller, A., Kronik, L., Neaton J. B. Quasiparticle and optical spectroscopy of the organic semiconductors pentacene and PTCDA from first principles. *Phys. Rev. B* **85,** 125307 (2012).
15. Sharifzadeh, S., Darancet, P., Kronik, L., Neaton, J. B. Low-energy charge-transfer excitons in organic solids from first-principles: the case of pentacene. *J. Phys. Chem. Lett.* **4,** 2197–2201 (2013).
16. Delgado M.C.R. et *al.* Impact of Perfluorination on the Charge-Transport Parameters of Oligoacene Crystals, *J. Am. Chem. Soc.* **131,** 1502-1512 (2009).
17. Smith, M. B.; Michl, J. Singlet Fission. *Chem. Rev.* **110,** 6891–6936 (2010).
18. Smith, M. B.; Michl, J. Recent Advances in Singlet Fission. *Annu. Rev. Phys. Chem.* **64,** 361–386 (2013).
19. Shockley, W., Queisser, H. J. Detailed balance limit of efficiency of pn junction solar cells. *J. Appl. Phys.* **32,** 510-519 (1961).
20. Congreve, D. N. et al. External quantum efficiency above 100% in a singlet-exciton-fission-based organic photovoltaic cell. *Science* **340**, 334-337 (2013).
21. Yanga, Y., Davidson, E. R., Yang, W. Nature of ground and electronic excited states of higher acenes. *Proc. Natl. Acad. Sci. USA* **113,** E5098-E5107 (2016)
22. Zimmerman, P. M., Zhang, Z., Musgrave, C. B. Singlet Fission in Pentacene through Multi-Exciton Quantum States. *Nature Chem.* **2,** 648–652 (2010).
23. Chan, W.-L. *et al*. Observing the Multiexciton State in Singlet Fission and Ensuing Ultrafast Multielectron Transfer. *Science* **334,** 1541–1545 (2011).
24. Busby, E. *et al*. Multiphonon relaxation slows singlet fission in crystalline hexacene. *J. Am. Chem. Soc.* **136,** 10654-10660 (2014).
25. Chernikov, A. *et al.* Spectroscopic Study of Anisotropic Excitons in Single Crystal Hexacene. *Phys. Chem. Lett.* **5,** 3632-3635 (2014).
26. Brütting, W. Physics of Organic Semiconductors. Wiley-VCH, Heidelberg, Germany (2005)

27. Sham, L. J., Rice, T. M. Many-particle derivation of the effective-mass equation for the Wannier exciton. *Phys. Rev.* **144**, 708-714 (1966).

28. Hanke, W., Sham, L. J. Local-field and excitonic effects in the optical spectrum of a covalent crystal. *Phys. Rev. B*, **12**, 4501-4511 (1975).

29. Sagmeister, S., Ambrosch-Draxl, C. Time-dependent density functional theory versus Bethe–Salpeter equation: an all-electron study. *Chem. Phys. Phys. Chem.* **11,** 4451-4457 (2009).

30. Runge, E., Gross, E. K. U. Density-functional theory for time-dependent systems. *Phys. Rev. Lett.* **52**, 997-1000 (1984).

31. Ghosh, S., Li, X. Q., Stepanenko, V., Würthner, F. Control of H- and J-Type pi Stacking by Peripheral Alkyl Chains and Self-Sorting Phenomena in Perylene Bisimide Homo- and Heteroaggregates. *Chem. Eur. J.* **14,** 11343-11357 (2008).

32. Sakamoto, Y. et al. Perfluoropentacene: high-performance p–n junctions and complementary circuits with pentacene. *J. Am. Chem. Soc.* **126,** 8138-8140 (2004).

33. Breuer, T., Witte, G. Epitaxial growth of perfluoropentacene films with predefined molecular orientation: a route for single-crystal optical studies. *Phys. Rev. B* **83,** 155428 (2011).

34. Hinderhofer, A. *et al.*, Optical properties of pentacene and perfluoropentacene thin films. *J. Chem. Phys.* **127,** 194705 (2007).

35. Spano, F.C. The spectral signatures of Frenkel polarons in H- and J-aggregates. *Acc. Chem. Res.* **43,** 429-439 (2010).

36. Huang, K., Rhys. A., Theory of light absorption and non-radiative transitions in F-centres. *Proc. Roy. Soc. A* **204,** 406-423 (1950).

37. Spanco, F. C. Theory of strong-field pump-probe spectroscopy in J-aggregates. *Chem. Phys. Lett.* **220,** 365-370 (1994).

38. Kolata, K., Breuer, T., Witte, G., Chatterjee, S. Molecular packing determines singlet exciton fission in organic semiconductors. *ACS nano* **8,** 7377–7383 (2014).

39. Jankowiak, H.-C., Stuber, J. L., Berger, R. Vibronic transitions in large molecular systems: Rigorous prescreening conditions for Franck-Condon factors. *J. Phys. Chem.* **127,** 23 (2007).

40. Berger, R. Fischer, C., Klessinger, M. Calculation of the vibronic fine structure in electronic spectra at higher temperatures. 1. benzene and pyrazine. *Journ. Phys. Chem. A* **102,** 7157-7167 (1998).

41. Huh, J., Berger, R. Coherent state-based generating function approach for Franck–Condon transitions and beyond. *J. Phys.: Conf. Ser.* **380,** 012019 (2012).

42. Huh, J., Neff, M., Rauhut, G., Berger, R. Franck–Condon profiles in photodetachment-photoelectron spectra of and based on vibrational configuration interaction wavefunctions. *Mol. Phys.* **108,** 409-423 (2010).

43. Treutler, O., Ahlrichs, R. Efficient Molecular Numerical Integration Schemes. *J. Chem. Phys.* **102,** 346 (1995).

44. Furche, F., Rappoport, D. *Density functional theory for excited states: equilibrium structure and electronic spectra.* Ch. III of "Computational Photochemistry", Ed. by M. Olivucci, Vol. 16 of "Computational and Theoretical Chemistry", Elsevier, Amsterdam (2005).

45. Weigend, F., Ahlrichs, R. Balanced basis sets of split valence, triple zeta valence and quadruple zeta valence quality for H to Rn: Design and assessment of accuracy. Phys. Chem. Chem. Phys. 7, 3297-305 (2005).

46. Weigend, F. Accurate Coulomb-fitting basis sets for H to Rn. Phys. Chem. Chem. Phys. 8, 1057-1065 (2006).

47. Becke, A. D. Density-functional thermochemistry. III. The role of exact exchange. *J. Chem. Phys.* **98,** 5648-5652 (1993).

48. Lee, C., Yang, W., Parr, R.G. Development of the Colle-Salvetti correlation-energy formula into a functional of the electron density. *Phys. Rev. B* **37,** 785-789 (1988).

49. TURBOMOLE V7.0 2015, a development of University of Karlsruhe and Forschungszentrum Karlsruhe GmbH, 1989-2007, TURBOMOLE GmbH, since 2007; available from http://www.turbomole.com.
50. Fulton, R. L. & Gouterman, M. Vibronic Coupling. I. Mathematical Treatment for Two Electronic States J. Chem. Phys. 35, 1059-1071 (1961).
51. Fulton, R. L., Gouterman, M. Vibronic Coupling. II. Spectra of Dimers. *J. Chem. Phys.* **41,** 2280-2286 (1964).
52. Schröter et al. Exciton-vibrational coupling in the dynamics and spectroscopy of Frenkel excitons in molecular aggregates. *Phys. Repots* **567,** 1-78 (2015).
53. Holstein, Davydov, Spano, F. C. Absorption and emission in oligo-phenylene vinylene nanoaggregates: The role of disorder and structural defects J. Chem. Phys., 2002, 116, 5877-5891